\documentclass[aip,reprint]{revtex4-1}

\usepackage{color}
\usepackage{amsmath}
\usepackage{amsfonts}
\usepackage{graphicx}

\begin{document}

\title{Quantum Transport With Two Interacting Conduction Channels}
\author{Alexander J. White}
\affiliation{Department of Chemistry \& Biochemistry, University of California San Diego, La Jolla, CA 92093, USA}
\author{Agostino Migliore}
\altaffiliation[Present address: ]{Department of Chemistry, Duke University, Durham, NC 27708, USA}
\affiliation{School of Chemistry, Tel Aviv University, Tel Aviv, 69978, Israel}
\author{Michael Galperin}
\affiliation{Department of Chemistry \& Biochemistry, University of California San Diego, La Jolla, CA 92093, USA}
\author{Abraham Nitzan}
\affiliation{School of Chemistry, Tel Aviv University, Tel Aviv, 69978, Israel}

\date{\today}

\begin{abstract}
The transport properties of a conduction junction model characterized by two mutually coupled 
channels that strongly differ in their couplings to the leads are investigated. 
Models of this type describe molecular redox junctions (where a level that is weakly coupled to 
the leads controls the molecular charge, while a strongly coupled one dominates the molecular 
conduction), and electron counting devices in which the current in a point contact is sensitive 
to the charging state of a nearby quantum dot. Here we consider the case where transport in 
the strongly coupled channel has to be described quantum mechanically 
(covering the full range between sequential tunneling and co-tunneling), 
while conduction through the weakly coupled channel is a sequential process that could by itself 
be described by a simple master equation. We compare the result of a full quantum calculation 
based on the pseudoparticle non-equilibrium Green function method to that obtained from 
an approximate mixed quantum-classical calculation, where correlations between the channels 
are taken into account through either the averaged rates or the averaged energy. 
We find, for the steady state current, that the approximation based on the averaged rates works 
well in most of the voltage regime, with marked deviations from the full quantum results only at the 
threshold for charging the weekly coupled level. These deviations are important for accurate 
description of the negative differential conduction behavior that often characterizes redox 
molecular junctions in the neighborhood of this threshold.
\end{abstract}




\maketitle

\section{Introduction}\label{intro}
Transport in mesoscopic and nanoscopic junctions is usually a multichannel phenomenon. 
Model studies of transport in junctions that comprise two, often interacting, conduction channels 
have been carried out in order to describe the essential features of different physical phenomena. 
Prominent examples are studies of interference effects in quantum conduction, analysis of single 
electron counting, where a highly transmitting junction (a point contact) is used to monitor 
the electronic state of a poorly transmitting one, and redox molecular junctions, where (transient) 
electron localization in one channel, stabilized by environmental polarization, determines the transition 
between redox states that are observed by the conduction properties of another channel. 
These three classes of phenomena are described by different flavors of the two-channel model. 
Interference is usually discussed as a single electron problem and interaction with the environment 
is minimized (often disregarded in model studies) so as to maintain phase coherent transport. 
Single electron counting with a point-contact detector  is by definition a many electron problem, 
however environmental interactions are again minimized (and again often disregarded in 
theoretical analysis) by lowering the experimental temperature in order to obtain detectable signals. 
Conduction in redox junctions is usually observed in room temperature polar environments and is 
characterized by large solvent reorganization that accompanies the electron localization at the redox site. 

\begin{figure}[t]
\centering\includegraphics[width=\linewidth]{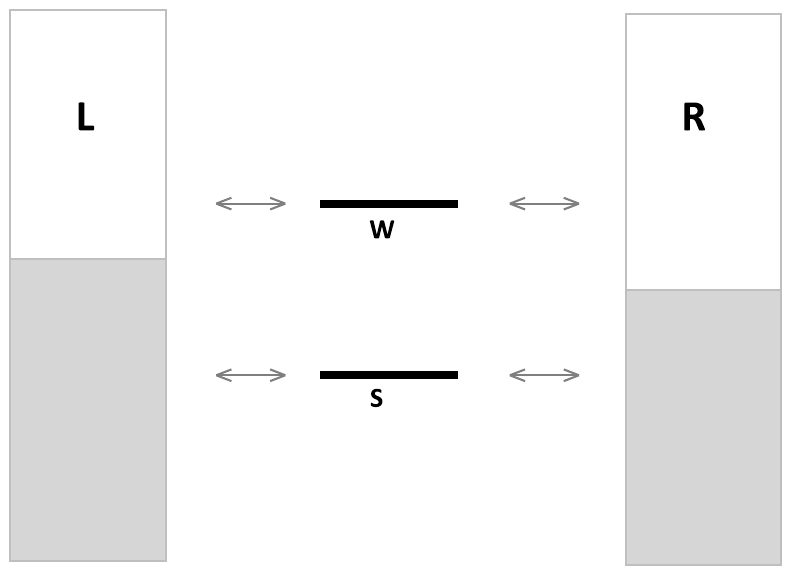}
\caption{\label{fig1}
The two channel model discussed in the paper. Each channel comprises 
one level coupled to the left and right electrodes. $W$ and $S$ denote weakly 
and strongly coupled levels, respectively.
}
\end{figure}

In recent work\cite{1,2,3,4}
we have studied the conduction properties of junctions of the latter type. 
We first analyzed, for a model involving a single conduction channel, 
the consequence of large solvent reorganization in the limit 
where the coupling between the molecular bridge and the metal leads 
is large relative to the frequency of the phonon mode used to model 
the solvent dynamical response.\cite{1,2,3} It was shown (using a mean field description 
essentially equivalent to the Born Oppenheimer approximation) 
that solvent induced stabilization of different charging states of 
the molecule can result in multistable operation of the junction, 
offering a possible rationalization of observations of negative 
differential resistance (NDR) and hysteretic response in molecular 
redox junctions. Such multistability was indeed observed recently 
in numerical simulations that avoid the mean field approximation.\cite{5,6}
Many redox junctions, however, operate in the opposite limit of 
relatively small molecule-lead coupling, where a single conduction channel 
model cannot show multistable transport behavior. Two of us have recently 
advanced a two channel model that can account for such observations.\cite{4}
In the absence of electron-phonon interaction (solvent polarization) this model is given  
by the Hamiltonian (see Fig.~\ref{fig1})
\begin{align}
\label{H}
\hat H =& \sum_{m=S,W}\varepsilon_m\hat d_m^\dagger\hat d_m
 + U\hat n_S\hat n_W + \sum_{k\in L,R}\varepsilon_k\hat c_k^\dagger\hat c_k
\nonumber \\
 +& \sum_{k\in L,R}\left(V_{kW}\hat c_k^\dagger\hat d_W+H.c.\right)
 \\
 +&\sum_{k\in L,R} \left(V_{kS}\hat c_k^\dagger\hat d_S+H.c.\right)
 \nonumber
\end{align}
where $\hat d_m^\dagger$ ($\hat c_k^\dagger$) creates electron in level 
$m$ (state $k$ of the contact), and $\hat n_m=\hat d_m^\dagger\hat d_m$,
$m=S,W$.
In this model, the two channels are coupled only capacitively (no inter-channel electron transfer). 
$U$ represents the standard Coulomb interaction between them.
Two coupled channel models such as (\ref{H}) also characterize single electron counting 
devices,\cite{8,9,10,11,12,13} where the current in a point contact 
(that can be represented by channel $S$) measures the charging state of a quantum dot 
used as a bridge in a nearby junction (channel $W$). The noise properties of such junctions 
have been studied extensively.\cite{14,15,16,17,18,19}

In this model, supplemented by electron phonon coupling that represents the response of 
a polar environment to the electronic occupations in levels $W$ and $S$,
the molecular redox site dominates the properties of 
one channel (addressed below as ``weakly coupled'' or ``slow'' and 
denoted by $W$), characterized by strong transient localization stabilized 
by large reorganization of the polar environment and weak coupling 
to the metal leads. Transport through this channel, that is, 
charging and discharging of the molecular redox site, was described 
by sequential kinetic processes.  A second channel (addressed below as 
``strongly coupled'' or ``fast'' and denoted by $S$) is more strongly coupled 
to the leads and is responsible for most or all of the observed current.\cite{20} 
Switching between charging states of the slow channel amounts to molecular 
redox states that affect the transmission, therefore the observed current, 
through the fast channel. Bistability and hysteretic response on 
experimentally relevant timescales are endowed into the model in a trivial 
way\cite{21} and, as was shown in Ref.~\onlinecite{4} (see also Refs.~\onlinecite{22,23,24}), 
NDR also appears naturally under suitable conditions.

Obviously, this behavior is generic and results from the timescale 
separation between the $W$ and $S$ channels together with the requirement 
that the observed current is dominated by the $S$ channel. 
In Refs.~\onlinecite{4} and \onlinecite{25}, 
we have described the expected phenomenology of such junction model 
in the limit where transport through both channels is described by simple 
kinetic equations with Marcus electron transfer rates. 
While, as indicated above, it is natural to model the slow dynamics 
(observed timescales $\sim 10^{-6}$~s) in this way, it is also of interest 
to consider fast channel transport on timescales where transport coherence 
is maintained. For example, one could envision a redox junction 
that switches between two conduction modes, which shows interference 
pattern associated with the structure of the fast channel. As a prelude 
for such considerations, we have studied in Ref.~\onlinecite{25}
also a model in which the weakly coupled channel $W$ is described 
by Marcus kinetics, however conduction through the strongly coupled 
channel $S$ is described as a coherent conduction process by means of the Landauer 
formula, assuming that the timescale of transport through this channel is 
fast enough to make it possible to ignore any interaction with the polar 
environment. As in any mixed quantum-classical dynamics, such description 
is not consistently derived from a system Hamiltonian, and ad-hoc assumptions 
about the way the quantum and classical subsystems interact with each other 
must be invoked, as described in Section~\ref{classical}.

In this paper we present a full quantum calculation of the current-voltage 
response of the two channel model described above, and use it to assess 
the approximate solution obtained using Eqs.~(\ref{rateEOM})-(\ref{nS}) 
with models A and B (see Section~\ref{classical}). The quantum calculation is done with 
the pseudoparticle non-equilibrium Green function (PP-NEGF) 
technique,\cite{ColemanPRB84,BickersRMP87,EcksteinWernerPRB10,OhAhnBubanjaPRB11}
named the slave boson technique when applied to a 3-states system 
(Anderson problem at infinite $U$),\cite{WingreenMeirPRB94,WingreenPRB96,HetlerKrohaHershfieldPRB98,KrohaWolflePRB00}
which was recently used by two of us to study effects of electron-phonon 
and exciton-plasmon interactions in molecular junctions.\cite{WhiteGalperinPCCP12,WhiteFainbergGalperinJPCL12}
We note that all the methods used in the paper have their own limitations. In particular, PP-NEGF is perturbative in the system-bath coupling. However, it accounts exactly for the intra-system interactions,
and it is the role of these interactions (quantum correlations due to system channels interactions) 
which is missed by the mixed quantum classical approaches and is the focus of the present study. 

In Section~\ref{classical} we present our model, briefly review the master equation description 
and introduce two approximate descriptions of mixed classical-quantum dynamics. 
The PP-NEGF technique and other details of the fully quantum calculation are described 
in Section~\ref{NEGF}. Section~\ref{numres} presents our results and discusses the validity 
of the approximate calculations. Section~\ref{conclude} concludes. 


\section{\label{classical}Mixed quantum classical approximations}
To account for the current-voltage behavior of a junction characterized by the Hamiltonian 
(\ref{H}), several workers\cite{14,15,16,17,18,19}
have used a master equation level of description, whereupon, for a given voltage, 
the dynamics of populating and de-populating the levels $S$ and $W$ is described by 
classical rate equations involving only their populations, with occupation and de-occupation 
rates given by standard expressions (see Eq.~(\ref{k}) below). Here, in order to focus on 
redox junction physics, the coupling of channel $W$ to the contacts is assumed to be much 
smaller than that of channel $S$, so that in the absence of correlations channel $W$ 
can be assumed to be classical and treated within such rate equations approach. 
At the same time channel $S$ will be treated as quantum, as discussed in the previous section.

 In Ref.~\onlinecite{25},
we have assumed that on the timescale of interest the junction can be in 
two states: $1$ and $0$, where the weakly coupled channel, that is 
the molecular redox site - is occupied or vacant, respectively. 
The probability $P_1=1-P_0$  that the junction is in state $1$ satisfies 
the kinetic equation
\begin{equation}
\label{rateEOM}
\frac{dP_1}{dt}=\left(1-P_1\right)k_{0\to 1}-P_1k_{1\to 0}
\end{equation}
where the rates $k_{0\to 1}$ and $k_{1\to 0}$ are electron transfer rates 
between a molecule and an electrode, here the rates to occupy and vacate
the redox molecular site, respectively. These rates are sums over contributions 
from the two electrodes
\begin{equation}
\label{kij}
k_{i\to j}=k_{i\to j}^{(L)}+k_{i\to j}^{(R)}; \qquad i,j=0,1
\end{equation}
and depend on the position of the redox molecular orbital energy
$\varepsilon_r$ relative to the Fermi energy (electronic chemical potential) 
of the corresponding electrode. In Ref.~\onlinecite{25}
we have used Marcus heterogeneous electron transfer theory to calculate these rates, 
thus taking explicitly into account solvent reorganization modeled as 
electron-phonon coupling in the high temperature and strong coupling limit. 
For the purpose of the present work it is enough to use the simpler, 
phonon-less, model
\begin{equation}
\label{k}
\begin{split}
 k_{0\to 1}^{(K)}\left(\varepsilon_r\right)=&
                 \Gamma_r^{K}f_K\left(\varepsilon_r\right)
 \\
 k_{1\to 0}^{(K)}\left(\varepsilon_r\right)=&
                 \Gamma_r^{K}\left[1-f_K\left(\varepsilon_r\right)\right]
\end{split}
\end{equation}
where $\varepsilon_r$ is the energy of the ``redox level'' (see below), 
$f_K(E)=\left[\exp\left((E-\mu_K)/T\right)+1\right]^{-1}$ ($K=L,R$)
is the Fermi-Dirac function of the electrode $K$, 
$\mu_K$ is the corresponding electronic chemical potential 
and $T$ is the temperature (in energy units).
$\Gamma_r^{K}$, $K=L,R$ are the widths of the redox molecular level 
due to its electron transfer coupling to the electrodes.\cite{36}
In terms of the Hamiltonian, Eq.~(\ref{H}) above, these widths are given by
$\Gamma_W^K=2\pi\sum_{k\in K}\lvert V_{Wk}\rvert^2\delta\left(E-\varepsilon_k\right)$. 
We have assumed that in the relevant energy regions these widths do not 
depend on energy.

From Eqs.~(\ref{rateEOM}) and (\ref{kij}), the steady state population 
of the redox site is $P_1=1-P_0=k_{0\to 1}/\left(k_{0\to 1}+k_{1\to 0}\right)$, 
and the current through the weakly coupled channel is
$I_W=k_{0\to 1}^{(L)}P_0-k_{1\to 0}^{(L)}P_1=k_{1\to 0}^{(R)}P_1-k_{0\to 1}^{(R)}P_0$. 
This current is however negligible relative to the contribution from 
the strongly coupled channel. In each of the states $0$ and $1$, 
the current $I_S$ as well as the average bridge population 
$\langle n_S\rangle$ in this channel, are assumed to be given by 
the standard Landauer theory for a channel comprising one single 
electron orbital of energy $\varepsilon_S$ bridging the leads, 
disregarding the effect of electron-phonon interaction,\cite{37,HaugJauho_2008}
\begin{align}
\label{IS}
&I_S\left(V;\varepsilon_S\right)=\frac{e}{\hbar}
\int_{-\infty}^{+\infty} \frac{d\varepsilon}{2\pi}\,
\frac{\Gamma_S^L\Gamma_S^R\left[f_L(E)-f_R(E)\right]}
{\left(\varepsilon-\varepsilon_S\right)^2+\left(\Gamma_S/2\right)^2}
\\
\label{nS}
&\langle n_S\left( V;\varepsilon_S\right)\rangle=
\int_{-\infty}^{+\infty}\frac{d\varepsilon}{2\pi}\,
\frac{\Gamma_S^Lf_L(\varepsilon)+\Gamma_S^Rf_R(\varepsilon)}
{\left(\varepsilon-\varepsilon_S\right)^2+\left(\Gamma_S/2\right)^2}
\end{align} 
where $\Gamma_S=\Gamma_S^L+\Gamma_S^R$ and where $\varepsilon_S$ and
$\Gamma_S^K$ take the values $\varepsilon_S^{(0)}$, $\Gamma_S^{K(0)}$
in state $0$, and $\varepsilon_S^{(1)}=\varepsilon_S^{(0)}+U$,
$\Gamma_S^{K(1)}=\Gamma_S^{K(0)}$ in state $1$.
$U$ is essentially a Coulomb energy term 
that measures the effect of electron occupation in channel $W$, 
i.e. at the redox site, on the energy of the bridging orbital in channel $S$.
$\Gamma_S^L$, $\Gamma_S^R$, $\varepsilon_S$, and $U$ 
are model 
parameters. The average population and current in channel S are given by
$\langle n_S\rangle=P_0\langle n_S\rangle^{(0)}+P_1\langle n_S\rangle^{(1)}$;
$\langle I_S\rangle=P_0 I_S^{(0)}+P_1 I_S^{(1)}$, 
where $I_S^{(0)}$ ($\langle n_S\rangle^{(0)}$) and 
$I_S^{(1)}$ ($\langle n_S\rangle^{(1)}$) are the values of $I_S$, Eq.~(\ref{IS})
($\langle n_S\rangle$, Eq.~(\ref{nS})) in system states $0$ (redox level empty), 
and $1$ (redox level populated).
Finally, the total current at a given voltage is $I=I_S+I_W\approx I_S$.

It should be noted that the rates defined by Eq.~(\ref{k}) are not completely 
specified, because the ``redox energy level'' $\varepsilon_r$ is not known: 
it is equal to $\varepsilon_W$ only if the capacitive interaction between 
the $S$ and $W$ channels is disregarded. To take this interaction into account, 
two models were examined in Ref.~\onlinecite{25}:\\
{\em Model A.} The rates are written as weighted averages over the populations 
$0$ and $1$ of channel $S$ with respective weights $1-\langle n_S\rangle$
and $\langle n_S\rangle$:
\begin{equation}
\begin{split}
 k_{0\to 1} =& \left(1-\langle n_S\rangle^{(0)}\right)k_{0\to 1}^{(S0)}
            + \langle n_S\rangle^{(0)}k_{1\to 0}^{(S0)}
 \\
 k_{1\to 0} =& \left(1-\langle n_S\rangle^{(1)}\right)k_{0\to 1}^{(S1)}
            + \langle n_S\rangle^{(1)}k_{1\to 0}^{(S1)}
\end{split}
\end{equation}
where $k_{0\to 1}^{(S0)}$, $k_{0\to 1}^{(S1)}$ are the rates to occupy 
and vacate, respectively, the redox site when the fast channel is not 
occupied, while $k_{0\to 1}^{(S1)}$, $k_{1\to 0}^{(S1)}$ 
are the corresponding rates when this channel is occupied. 
The dependence of these rates on the occupation of the fast channel is 
derived from the dependence of $\varepsilon_r$ in Eq.~(\ref{k}) 
on the occupation of level $S$: $\varepsilon_r=\varepsilon_W$   
when this level is not occupied, and $\varepsilon_r=\varepsilon_W+U$
when it is. That is,
\begin{equation}
\begin{split}
k_{0\to 1}^{(K,S0)}=& \Gamma_r^Kf_K(\varepsilon_W) \\
k_{1\to 0}^{(K,S0)}=& \Gamma_r^K\left[1-f_K(\varepsilon_W)\right]
\\
k_{0\to 1}^{(K,S1)}=& \Gamma_r^Kf_K(\varepsilon_W+U) \\
k_{1\to 0}^{(K,S1)}=& \Gamma_r^K\left[1-f_K(\varepsilon_W+U)\right]
\end{split}
\end{equation}
Here $K=L,R$.\\
{\em Model B.} The rates are given by Eq.~(\ref{k}), 
with $\varepsilon_r$ calculated as the difference between the energies of 
two molecular states, one with the redox level populated,
$E_1=\left(\varepsilon_S^{(0)}+U\right)\langle n_S\rangle^{(1)}+\varepsilon_2^{(0)}=\varepsilon_S^{(1)}\langle n_S\rangle^{(1)}+\varepsilon_2^{(0)}$ 
and the other with the redox level empty,
$E_0=\varepsilon_S^{(0)}\langle n_S\rangle^{(0)}$:
\begin{equation}
\varepsilon_r=
\left(\varepsilon_S^{(1)}\langle n_S\rangle^{(1)}+\varepsilon_2^{(0)}\right)
-\varepsilon_S^{(0)}\langle n_S\rangle^{(0)}\\
\end{equation}

These two models are associated with different physical pictures that 
reflect different assumptions about relative characteristic timescales. 
Model A assumes that the switching rates between states $0$ and $1$ follow 
the instantaneous population in channel $S$, while model B assumes that these 
switching rates are sensitive only to the average population
$\langle n_S\rangle$. Model B results from a standard Hartree approximation 
that would be valid if the electronic dynamics in channel $W$ is slow relative 
to that of channel $S$ (see Appendix). From the discussion above it may appear at first glance 
to be the case, since transmission through channel $W$ is small, 
implying that the rates $k_{0\to 1}$ and $k_{1\to 0}$ are small. 
However, the electronic process that determines the timescale on which 
these rates change is not determined by the magnitude of these rates but by
the response of the electrodes to changes in
$\varepsilon_r$ following changes in the bridge level population of 
the strongly coupled channel. This characteristic time (or times), $\tau_B$, which is bounded 
below by the inverse  electrode bandwidth, may depend also on temperature and the energy
dependence of the spectral density, and can be shorter than the timescale of 
order of $\Gamma_S^{-1}$ on which population in channel $S$ is changing
(note that $\tau_B$ is vanishingly short in the wide band limit). 
In this case model A would provide a better approximation. 
For comparison, we also present below results for model C, in which 
the effect of the interaction between the two channels on the electron 
transfer kinetics in channel $W$ is disregarded so that
\begin{equation}
\begin{split}
 k_{0\to 1}^K=&\Gamma_W^K f_K(\varepsilon_W)
 \\
 k_{1\to 0}^K=&\Gamma_W^K\left[1-f_K(\varepsilon_W)\right]
 \end{split}
\end{equation}
while the current through channel $S$ continues to be sensitive to 
the difference between states $0$ and $1$, as before.

\begin{figure}[t]
\centering\includegraphics[width=\linewidth]{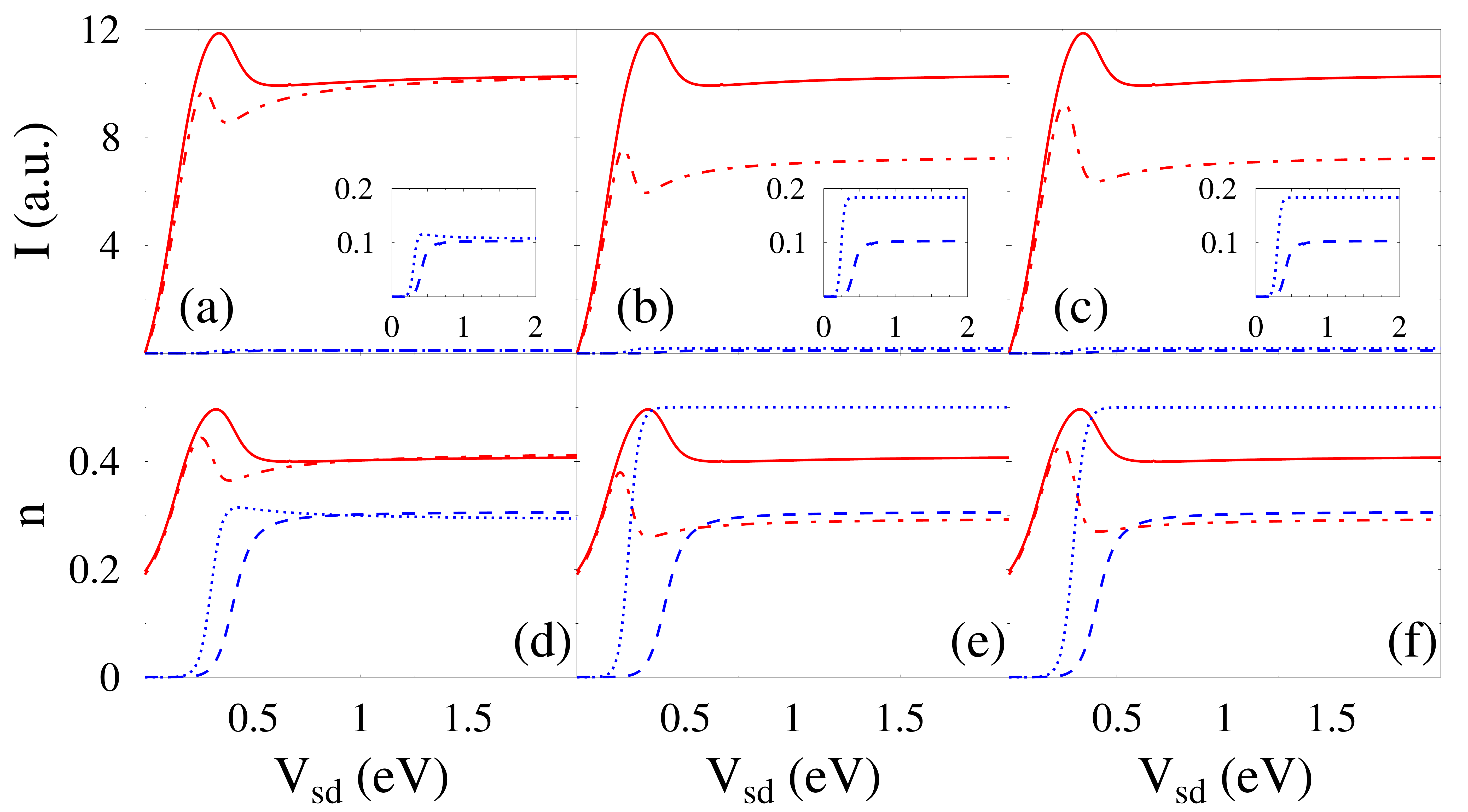}
\caption{\label{fig2}
(Color online) Current (panels a-c) and populations of the channels (panels d-f).
Results for the models A (panels a and d), B (panels b and e), and C (panels c and f)
are shown for the channels S (dash-dotted line, red) and W (dotted line, blue),
and compared to the PP-NEGF results for the same channels (solid, red and dashed, blue lines, respectively). Note, the PP-NEGF data is the same in panels a-c and d-f.
See text for parameters.
}
\end{figure}

\section{\label{NEGF}The pseudoparticle Green function method}
Models A and B above represent attempts to partly account for the coupling between 
channels within the classical rate equations description of channel $W$. 
The existence of capacitive coupling between the channels makes such mixed 
quantum-classical description potentially invalid, since it misses quantum correlations between 
the two channels. To estimate the performance of these approximations we shall compare them 
to a fully quantum calculation based on the pseudoparticle nonequilibrium Green 
function technique.\cite{WhiteGalperinPCCP12}

In the PP-NEGF approach, a set of molecular many-body states, 
$\{\lvert N\rangle\}$, defines the set of pseudoparticles to be considered, 
so that one pseudoparticle represents each state. In particular, for the model 
(\ref{H}) the molecular subspace of the problem is represented by four 
many-body states: $\lvert N\rangle=\lvert n_W,n_S\rangle$, 
where $n_{W,S}=0,1$. Let $\hat p_N^\dagger$ ($\hat p_N$) be the creation 
(annihilation) operator for the state $\lvert N\rangle$. 
These operators are assumed to satisfy the usual fermion or boson commutation 
relations depending on the type of the state. In our case the pseudoparticles 
associated with the states $\lvert 1,0\rangle$ and $\lvert 0,1\rangle$ 
are of Fermi type, while those corresponding to states $\lvert 0,0\rangle$ 
and $\lvert 1,1\rangle$ follow Bose statistics. 
The PP-NEGF is defined on the Keldysh contour as  
\begin{equation}
 G_{N_1,N_2}(\tau_1,\tau_2) \equiv -i\langle T_c\, \hat p_{N_1}(\tau_1)\,
 \hat p_{N_2}^\dagger(\tau_2)\rangle
\end{equation}
In the extended Hilbert space it satisfies the usual Dyson equation, 
thereby providing a standard machinery for their evaluation. 
Reduction to the physically relevant subspace of the total pseudoparticle 
Hilbert space is achieved by imposing the constraint 
\begin{equation}
 \sum_N \hat p_N^\dagger\hat p_N =1
\end{equation}
on the Dyson equation projections. The resulting system of equations 
for the Green function projections has to be solved self-consistently 
(see e.g. Ref.~\onlinecite{WhiteGalperinPCCP12} for details). 
Finally, connections to Green functions of the standard NEGF formulation 
can be obtained by using relations between the electron operators in 
the molecular subspace of Eq.~(\ref{H}) and those of the pseudoparticles 
\begin{equation}
 \hat d_m^\dagger = \sum_{N_1,N_2}
 \langle N_1\rvert\hat d_m^\dagger\lvert N_2\rangle 
 \hat p_{N_1}^\dagger \hat p_{N_2}
\end{equation}
Thus the current through the junction can be obtained either by the usual 
NEGF expression,\cite{HaugJauho_2008} or within its pseudoparticle 
analog.\cite{WhiteGalperinPCCP12}

Results of calculations based on this procedure and on the kinetic schemes 
described in Section~\ref{intro} are presented and discussed next. 


\begin{figure}[t]
\centering\includegraphics[width=\linewidth]{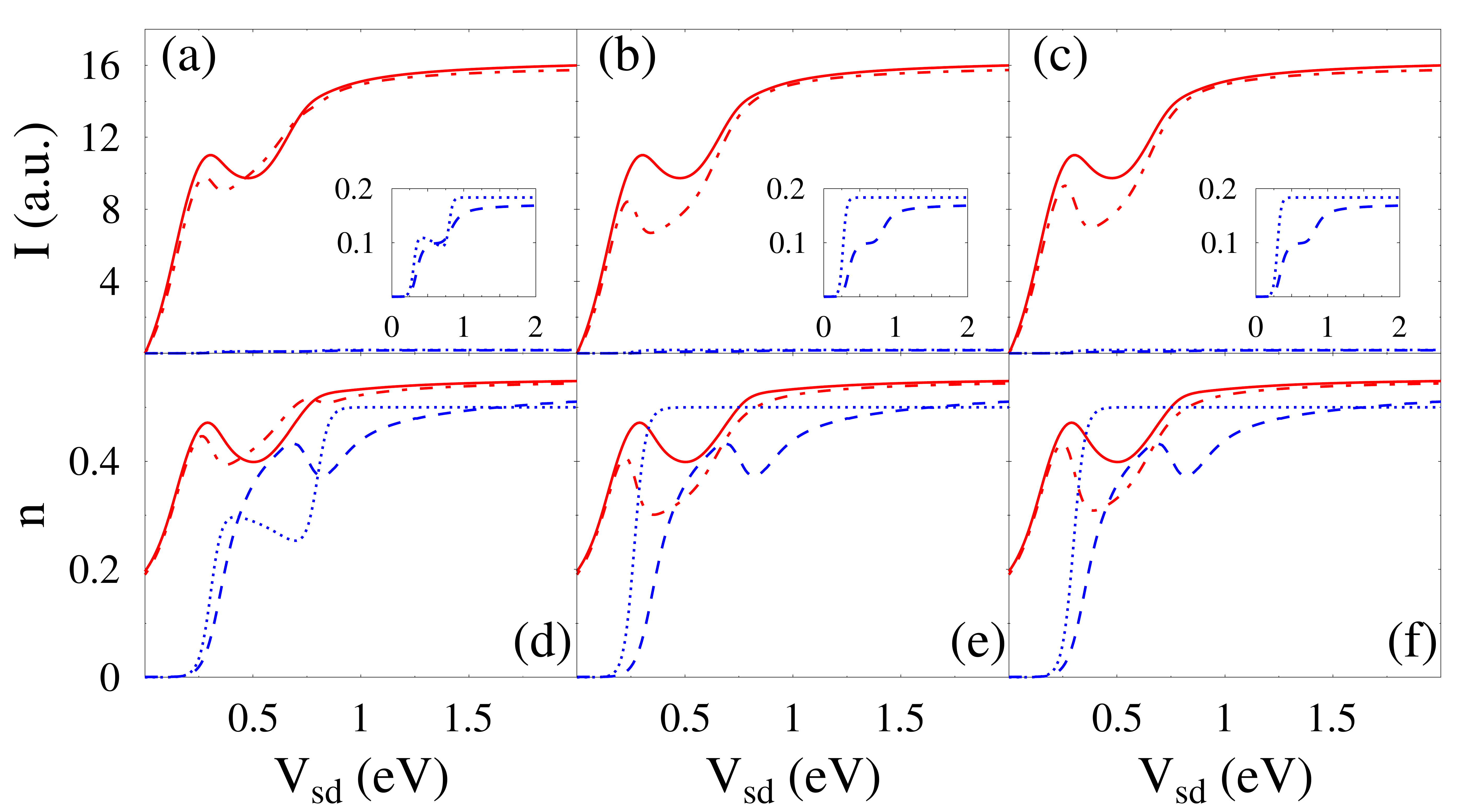}
\caption{\label{fig3} 
(Color online) Same as Fig.~\ref{fig2} except $U=500$~meV
}
\end{figure}

\section{\label{numres}Results and discussion}
\begin{figure}[t]
\centering\includegraphics[width=\linewidth]{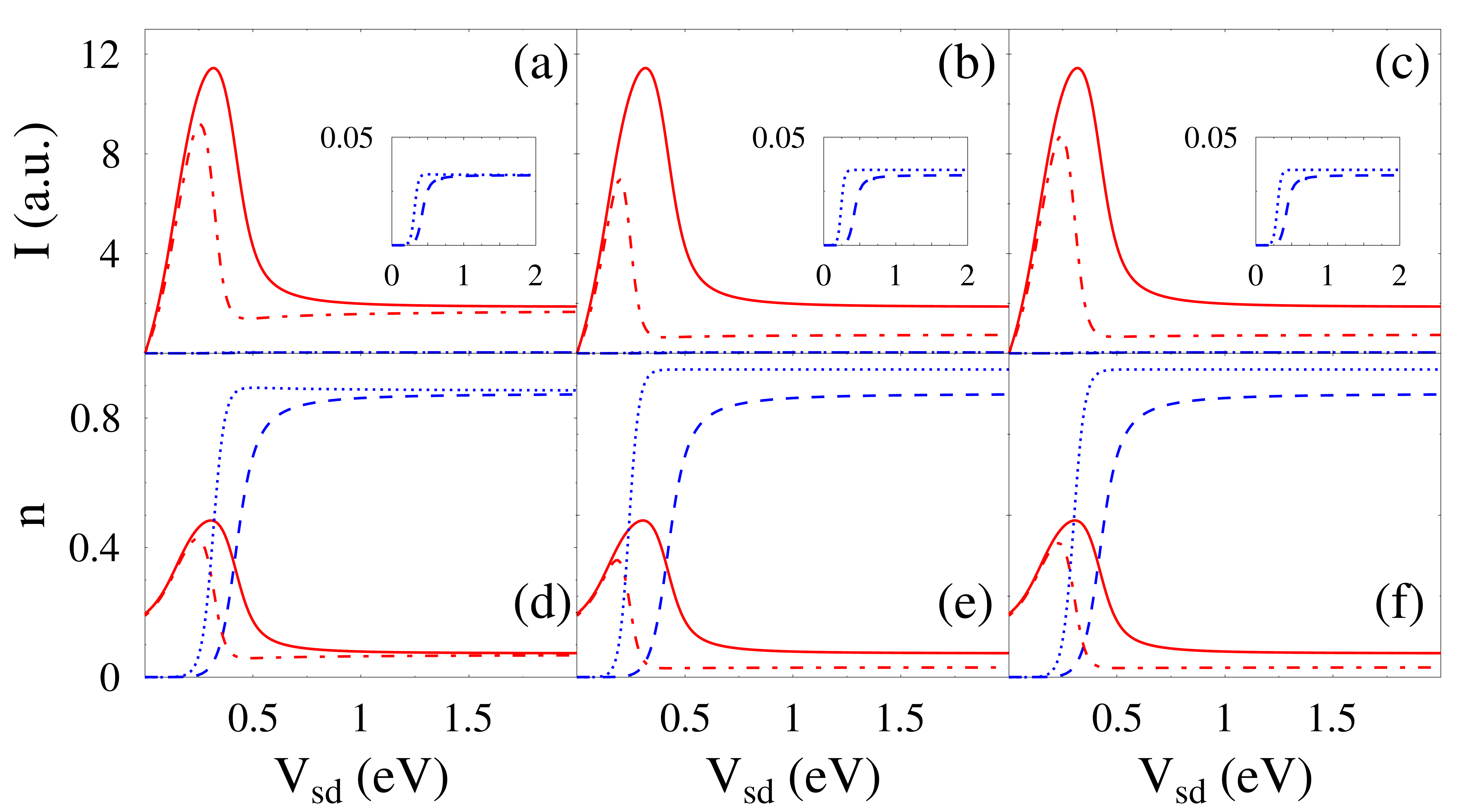}
\caption{\label{fig4}
(Color online) Same as Fig.~\ref{fig2} except $\Gamma_W^L=1.9$~meV and $\Gamma_W^R=0.1$~meV.
}
\end{figure}

\begin{figure}[t]
\centering\includegraphics[width=\linewidth]{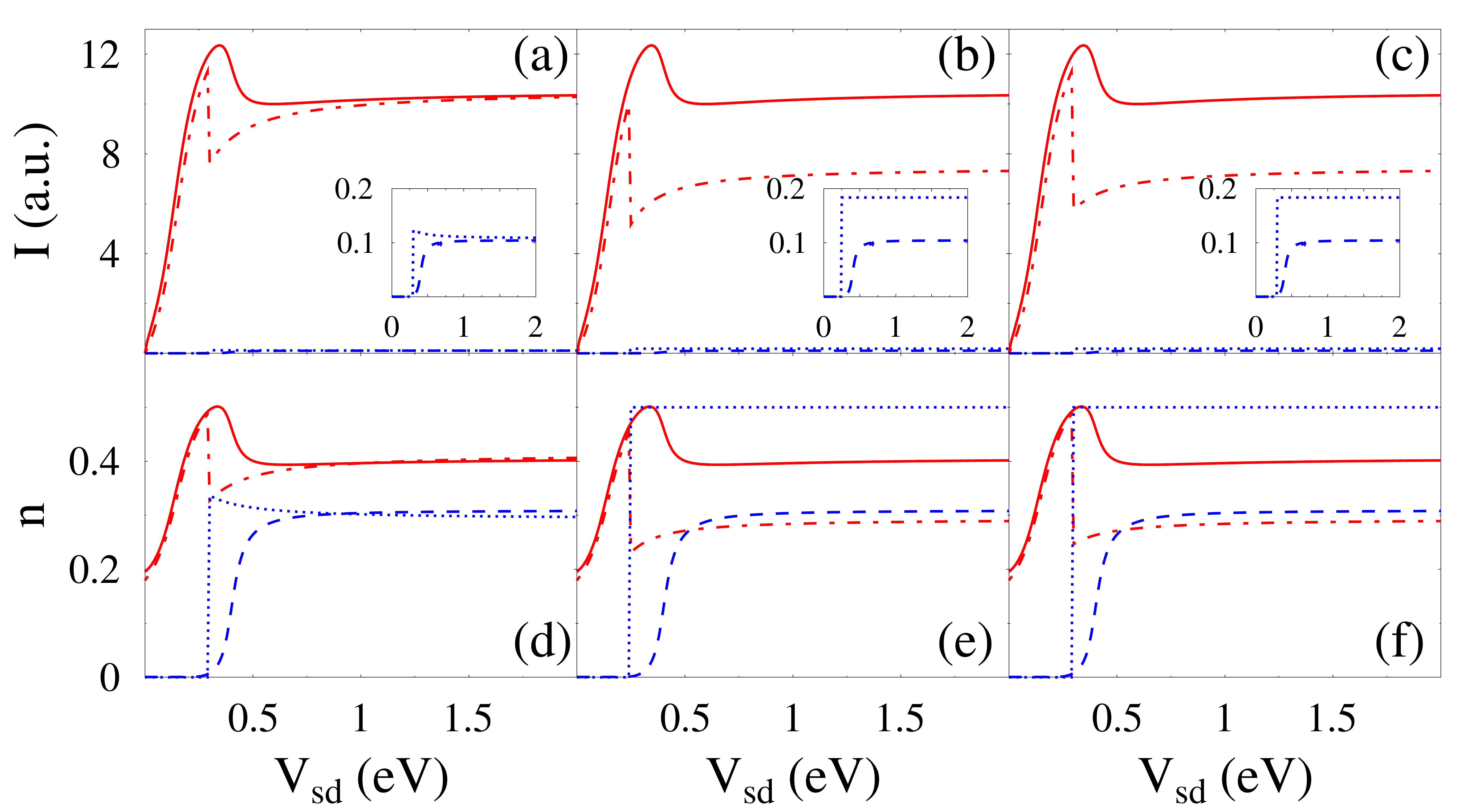}
\caption{\label{fig5}
(Color online) Same as Fig.~\ref{fig2} except $T=0$~K. 
}
\end{figure}

In Figures~\ref{fig2} and \ref{fig3} we compare results from the fully 
quantum calculation based on the PP-NEGF technique with those based on 
the kinetic approximations defined by models A-C of Section~\ref{intro}. 
Panels (a), (b) and (c) in Fig.~\ref{fig2} show the current through channels 
$S$ (red) and $W$ (blue) as function of voltage, 
while the corresponding panels (d), (e) and (f) show, with the same color 
and line-forms codes, the electronic populations in these channels. 
The full and dashed lines in these plots correspond to the PP-NEGF 
calculations for channels $S$ and $W$, respectively, and are identical in 
the panels (a-c) and in panels (d-f). The dash-dotted and dotted lines 
show results based on models A (panels (a) and (d)), B (panels (b) and (e)) 
and C (panels (c) and (f)).   
The parameters used in these calculations are $E_F=0$, $T=300$~K,
$\Gamma^L_W=\Gamma^R_W=1$~meV, $\Gamma^L_S=\Gamma^R_S=100$~meV,
$\varepsilon_S=150$~meV, $\varepsilon_W=300$~meV, 
and $U=10$~eV. For this choice of $U$ states $S$ and $W$ cannot be populated 
simultaneously. The corresponding panels of Figs.~\ref{fig3} and \ref{fig4} 
show similar results for the same choice of parameters, except 
that in Fig.~\ref{fig3} $U$ is taken $500$~meV while in Fig.~\ref{fig4} 
$\Gamma_W^L=1.9$~meV and $\Gamma_W^R=0.1$~meV 
(so $\Gamma_W=\Gamma_W^L+\Gamma_W^R=2$~meV as before).
The latter choice designates level $W$ as a blocking level - current goes down considerably 
when the voltage bias exceeds the threshold ($300$~meV) needed to populate it), 
and has been suggested before\cite{4a,4b,4c,4}
as a model for negative differential resistance in molecular junctions. 
Finally, in Fig.~\ref{fig5}, the parameters are the same is in Fig.~\ref{fig2} except that
$T=0$~K. The voltage was changed by moving the Fermi level of 
the left electrode, keeping the right electrode static. 
The insets in the $I/V$ plots show a closeup look at the contribution from 
channel $W$. 
The following observations are notable:\\
(a) In comparison with the full quantum calculation, Model A performs 
considerably better than model B and, not surprisingly, than model C. 
The failure of model B is notable in view of the common practice 
to use the timescale separation as an argument for applying mean field theory 
in such calculations; however, as argued above, it follows from the use of the wide band limit
for the electrodes in the calculations.\\
(b) While model A seems to be quite successful in much of the voltage regime, 
it fails, as expected, near and around $V=0.3$~V, the (bare) threshold 
to populate the $W$ level. It is at this point of maximal fluctuations 
in the $W$ population that electronic correlation is most pronounced, 
as this population is strongly correlated with that in S.\\
(c) The deviation of the kinetic approximation from the full quantum result 
is considerably larger for the current and population of channel $W$ 
(the redox site) than for channel $S$. This reflects the fact that the rates 
of charging and discharging the redox site are sensitive to its correlation 
with the population on the strongly coupled level, while the dynamics of 
the latter responds most of the time just to the static population in $W$. 
Of course, these large deviations in the current carried by channel $W$ 
have only an insignificant effect on the overall observed current. 
To see these important quantum correlation effects one would need 
to monitor directly the electronic population of the redox site, 
which is possible in principle using spectroscopy probes.\\
(d) As a model for negative differential resistance (Fig.~\ref{fig4}), model A performs 
qualitatively well, however the full calculation sets the NDR threshold considerably 
higher than that predicated by the approximate calculation.\\
(e) As expected, the differences between the full quantum calculation and the results of 
model A become more pronounced at  $T=0$~K. While the results of model A display 
sharp threshold behavior, the full calculation is much less sensitive to temperature 
for the present choice of parameters because the width of the transition region 
is dominated by $\Gamma_S$ that is substantially greater than the thermal energy.


\section{\label{conclude}Conclusion}
We have examined the electronic transport behavior of a generic junction 
model that comprises a bridge characterized by two interacting transport 
channels whose couplings to the leads are vastly different from each other. 
This is a model for a molecular redox junction and also for a point contact 
detector interacting with a weakly coupled nanodot bridge. We have compared 
approximate kinetic schemes for the dynamics of this junction to a full 
quantum calculation based on the pseudoparticle NEGF methodology. 
We found that a kinetic model in which the electron transfer rates 
in the weakly coupled channel (redox site) respond instantaneously 
to occupation changes in the strongly coupled channel works relatively well 
in comparison with a mean field calculation. Still, this model fails quantitatively when 
the molecular level comes close to the electrochemical potential of the  lead, 
reflecting the significance of electronic correlations in this voltage range.

This paper has focused on the steady state current. Correlations between the two channels are 
expected to become considerably more pronounced in the noise properties of such junctions and, 
most probably, would not be amenable to analysis using the kinetic approximation of model A. 
We defer this interesting issue to future work.


\begin{acknowledgments}
The research of AN is supported by the Israel Science Foundation, 
the Israel-US Binational Science Foundation and the European Research 
Council under the European Union's Seventh Framework Program 
(FP7/2007-2013; ERC grant agreement no 226628). 
MG gratefully aknowledges support by the Department of Energy 
(Early Career Award, DE-SC0006422) and the US-Israel Binational Science 
Foundation (grant no. 2008282). 
We thank Kristen Kaasbjerg for useful discussions.
MG and AN  thank the KITPC Beijing for hospitality and support during 
the time when this work was completed.
\end{acknowledgments}

\appendix
\section{Timescale considerations leading to the models A and B}
When it is reasonable to speak about rate of a channel,
the formal expression for the $W$ channel rate is\ref{EspGalp2009}
\begin{equation}
\label{rate}
 \int_{-\infty}^{t} dt'\, e^{i\int_{t'}^tds\,\varepsilon_r(s)}
 V(t)\, C(t-t')\, V(t')
\end{equation}
where $\epsilon_r$ is the position of the redox level, $V(t)$ is the coupling between the channel $W$
and the bath, 
and $C(t-t')$ is the bath correlation time.

At least two timescales have to be taken into account: one related to
the dynamics of the redox level, $\varepsilon_r(t)$, the other
representing characteristic timescale of the bath. Note, that in general
the bath is characterized by several timescales
(e.g. the bandwidth of the metal, temperature, and variation of spectral density).
In our case the characteristic timescale for the dynamics of the level in the $W$
channel is given by the rate of population change in the $S$ channel.
The latter is proportional to $\Gamma_S^{-1}$ (Coulomb interaction is
instantaneous). Let assume that the characteristic
time of the bath is $\tau_B$. The two extremes are $\tau_B\ll\Gamma_S^{-1}$
and $\tau_B\gg\Gamma_S^{-1}$. The former case corresponds to
slow motion of the level relative to the bath dynamics, so that
expression (\ref{rate}) yields a set of rates ($2$ in our case) for
different positions of the redox level. This corresponds to the model $A$ of
the paper.

The other extreme,  $\tau_B\gg\Gamma_S^{-1}$, corresponds to quick motion of the
redox level position, which requires averaging of the exponential factor
in (\ref{rate}). This leads to appearance of a single rate, calculated
at the average position of the level, which is model $B$.



\begin{thebibliography}{99}
\bibitem{6}It should be emphasized that the theory addresses only locally stable states that will not persist beyond some finite lifetime, and do not imply multistability in the thermodynamic sense.
\bibitem{20}The ``fast'' channel carries all the current if the ``slow'' channel is coupled only to one of the leads.
\bibitem{36}Note that while in quantum mechanics damping rates and level widths are synonymous, it is exactly the energetic consequence of the finite lifetime, that is, the level broadening, which is disregarded in the kinetic approximation.
\end{thebibliography}
\end{document}